\documentclass{ws-mpla}
\begin{document}
\markboth{C. Zhang, H. Liu and L. Xu} {Universe Evolution in a
$5D$ Ricci-flat Cosmology}

%
\catchline{}{}{}{}{} %
\title{UNIVERSE EVOLUTION IN A $5D$ RICCI-FLAT COSMOLOGY}
\author{CHENGWU ZHANG, HONGYA LIU\thanks{%
hyliu@dlut.edu.cn} \,\,and LIXIN XU}
\address{Department of Physics, Dalian University of Technology, Dalian
116024, P.R. China}
\author{PAUL S. WESSON}
\address{Department of Physics, University of Waterloo,
Waterloo, Ontario \ N2L 3G1, Canada} \maketitle

\pub{Received (Day Month Year)}{Revised (Day Month Year)}

\begin{abstract}
We use Wetterich's parameterization equation of state (EOS) of dark energy
to a $5D$ Ricci-flat cosmological solution and we suppose the universe
contains three major components: matter, radiation and dark energy. By using
the relation between the scale factor and the redshift $z$, we show that the
two arbitrary functions contained in the $5D$ solution could be solved out
analytically in terms of the variable $z$. Thus the whole $5D$ solution
could be constructed uniquely if the current values of the three density
parameters $\Omega _{m0}$, ${\Omega _{r0}}$, $\Omega _{x0}$, the EOS $w_{0}$%
, and the bending parameter $b$ contained in the EOS are all known.
Furthermore, we find that all the evolutions of the mass density $\Omega _{m}
$, the radiation density ${\Omega _{r}}$, the dark energy density $\Omega
_{x}$, and the deceleration parameter $q$ depend on the bending parameter $b$
sensitively. Therefore it is deserved to study observational constraints on
the bending parameter $b$.
\end{abstract}


\ccode{PACS Nos.:04.50.+h, 98.80.-k, 98.80.Es}

\section{Introduction}

Observations of Cosmic Microwave Background (CMB) anisotropies
indicate that the universe is flat and the total energy density is
very close to the critical one with $\Omega _{total}\simeq 1$
\cite{CMB}. Meanwhile, observations of high redshift type Ia
supernovae\cite{Ia} reveal the speeding up expansion of our
universe and the surveys of clusters of galaxies show that the
density of matter is very much less than the critical
density\cite{SDSS}. These three tests nicely complement each other
and indicate that an exotic component with negative pressure
dubbed dark energy dominates the present universe. Various dark
energy models have been proposed among them the most promising
ones are probably those with a scalar field such as
quintessence\cite{Quintessence}, phantom\cite{Phantom}, K-essence\cite%
{K-essence} and so on. For this kind of models, one can design
many kinds of potentials\cite{VS} and then study EOS for the dark
energy. Another way is to use a parameterization of the EOS to fit
the observational data, and then to reconstruct the potential
and/or the evolution of the universe\cite{reconstructure}. The
latter has the advantage that it does not depend on a specified
model of the dark energy and, therefore, is also called a
model-independent method\cite{Corasaniti}.

Both the classical Kaluza-Klein theories and the modern
string/brane theories require the existence of extra dimensions.
If the universe has more than four dimensions, general relativity
should be extended from $4D$ to higher dimensions. One of such
extensions is the $5D$ Space-Time-Matter (STM)
theory\cite{Wesson,JMOverduin} in which our universe is a $4D$
hypersurface floating in a $5D$ Ricci-flat manifold. This theory
is supported by Campbell's theorem which states that any
analytical solution of the $ND$
Einstein equations can be embedded in a $(N+1)D$ Ricci-flat manifold\cite%
{Campbell}. A class of cosmological solutions of the STM theory is
given by Liu, Mashhoon and Wesson\cite{LMW,LandMashhoon}. It was
shown that dark energy models, similar as the 4D quintessence and
phantom ones, can also be constructed in this $5D$ cosmological
solution in which the scalar field is induced from the $5D$
vacuum\cite{ChangLiu,Liuetal}. The purpose of this paper is to use
a model-independent method to study the EOS of the dark energy and
the evolution of the $5D$ universe.

Various parameterization of the EOS of dark energy have been
presented and
investigated\cite{Corasaniti,Linder,Upadhye,Wang,David}. In this
paper we will study one of them presented firstly by
Wetterich\cite{Wetterich} where there is a bending parameter $b$
which describes the deviation of the EOS from a constant $w_{0}$.
The paper is organized as follows. In Section 2, we introduce the
$5D$ Ricci-flat cosmological solution and derive the densities for
the three major components of the universe. In Section 3, we will
reconstruct the evolution of the model with different values of
the bending parameter $b$. Section 4 is a short discussion.

\section{Density Parameters in the $5D$ Model}

The $5D$ cosmological solutions read\cite{LMW}
\begin{equation}
dS^{2}=B^{2}dt^{2}-A^{2}\left( \frac{dr^{2}}{1-kr^{2}}+r^{2}d\Omega
^{2}\right) -dy^{2},  \label{5-metric}
\end{equation}%
with $d\Omega ^{2}\equiv \left( d\theta ^{2}+\sin ^{2}\theta d\phi
^{2}\right) $ and
\begin{eqnarray}
A^{2} &=&\left( \mu ^{2}+k\right) y^{2}+2\nu y+\frac{\nu ^{2}+K}{\mu ^{2}+k},
\nonumber \\
B &=&\frac{1}{\mu }\frac{\partial A}{\partial t}\equiv \frac{\dot{A}}{\mu },
\label{A-B}
\end{eqnarray}%
where $\mu =\mu (t)$ and $\nu =\nu (t)$ are two arbitrary functions of $t$, $%
k$ is the $3D$ curvature index $\left( k=\pm 1,0\right) $, and $K$ is a
constant. This solution satisfies the $5D$ vacuum equations $R_{AB}=0$. So
we have
\begin{eqnarray}
I_{1} &\equiv &R=0,I_{2}\equiv R^{AB}R_{AB}=0,  \nonumber \\
I_{3} &=&R_{ABCD}R^{ABCD}=\frac{72K^{2}}{A^{8}},  \label{3-invar}
\end{eqnarray}%
which shows that $K$ determines the curvature of the $5D$ manifold. The
Hubble and deceleration parameters are\cite{Liu-b},
\begin{eqnarray}
H &\equiv &\frac{\dot{A}}{AB}=\frac{\mu }{A}  \label{H} \\
q\left( t,y\right)  &\equiv &\left. -A\frac{d^{2}A}{d\tau
^{2}}\right/ \left( \frac{dA}{d\tau }\right)
^{2}=-\frac{A\dot{\mu}}{\mu \dot{A}}. \label{q}
\end{eqnarray}

Using the $4D$ part of the $5D$ metric (\ref{5-metric}) to calculate the $4D$
Einstein tensor, one obtains
\begin{eqnarray}
^{(4)}G_{0}^{0} &=&\frac{3\left( \mu ^{2}+k\right) }{A^{2}},  \nonumber \\
^{(4)}G_{1}^{1} &=&^{(4)}G_{2}^{2}=^{(4)}G_{3}^{3}=\frac{2\mu \dot{\mu}}{A%
\dot{A}}+\frac{\mu ^{2}+k}{A^{2}}.  \label{einstein}
\end{eqnarray}%
So the $4D$ induced energy-momentum tensor can be defined as $T^{\alpha
\beta }=^{(4)}G^{\alpha \beta }$. In this paper we consider the case where
the $4D$ induced matter $T^{\alpha \beta }$ is composed of three components:
matter $\rho _{m}$, radiation $\rho _{r}$ and dark energy $\rho _{x}$, which
are minimally coupled to each other. So we have
\begin{eqnarray}
\frac{3\left( \mu ^{2}+k\right) }{A^{2}} &=&\rho _{m}+\rho _{r}+\rho _{x},
\nonumber \\
\frac{2\mu \dot{\mu}}{A\dot{A}}+\frac{\mu ^{2}+k}{A^{2}}
&=&-p_{m}-p_{r}-p_{x},  \label{FRW-Eq}
\end{eqnarray}%
with
\begin{eqnarray}
\rho _{m} &=&\rho _{m0}A_{0}^{3}A^{-3},\quad p_{m}=0,\quad \rho _{r}=\rho
_{r0}A_{0}^{4}A^{-4}=3p_{r},  \label{EOS-M} \\
p_{x} &=&w_{x}\rho _{x}.  \label{EOS-X}
\end{eqnarray}%
From Eqs. (\ref{FRW-Eq}) - (\ref{EOS-X}) and for $k=0$, we obtain
the EOS of the dark energy
\begin{equation}
w_{x}=\frac{p_{x}}{\rho _{x}}=-\frac{2\mu \dot{\mu}/\left( A\dot{A}\right)
+\mu ^{2}/A^{2}+\rho _{r0}A_{0}^{4}A^{-4}/3}{3\mu ^{2}A^{2}-\rho
_{m0}A_{0}^{3}A^{-3}-\rho _{r0}A^{-4}},  \label{wx}
\end{equation}%
and the dimensionless density parameters
\begin{eqnarray}
\Omega _{m} &=&\frac{\rho _{m}}{\rho _{m}+\rho _{r}+\rho _{x}}=\frac{\rho
_{m0}A_{0}^{3}}{3\mu ^{2}A},  \label{omiga-M} \\
\Omega _{r} &=&\frac{\rho _{r}}{\rho _{m}+\rho _{r}+\rho _{x}}=\frac{\rho
_{r0}A_{0}^{4}}{3\mu ^{2}A^{2}},  \label{omiga-R} \\
\Omega _{x} &=&1-\Omega _{m}-\Omega _{r}.  \label{omiga-X}
\end{eqnarray}%
where $\rho _{m0}$ and $\rho _{r0}$ are the current values of matter and
radiation densities, respectively.

Consider equation (\ref{wx}) where $A$ is a function of $t$ and $y$.
However, on a given $y=$ constant hypersurface, $A$ becomes $A=A(t)$. As
noticed before\cite{Xu-Recons}, the term $\dot{\mu}/\dot{A}$ in (\ref{wx})
can now be rewritten as $d\mu /dA$. Furthermore, we use the relation
\begin{equation}
A_{0}/A=1+z,  \label{factor}
\end{equation}%
and define $\mu _{0}^{2}/\mu ^{2}=f(z)$ (with $f(0)\equiv 1$), and
then we find that equations~(\ref{wx})-(\ref{omiga-X}) and
(\ref{q}) can be expressed in term of the redshift $z$ as
\begin{equation}
w_{x}=-\frac{1+\Omega _{r}+(1+z)dlnf(z)/dz}{3(1-\Omega _{m}-\Omega
_{r})}, \label{wxz}
\end{equation}

\begin{equation}
\Omega_m=\Omega_{m_0}(1+z)f(z),  \label{Omegamz}
\end{equation}

\begin{equation}
\Omega_r=\Omega_{r_0}(1+z)^2 f(z),  \label{Omegarz}
\end{equation}

\begin{equation}
\Omega_x=1-\Omega_m-\Omega_r,  \label{Omegaxz}
\end{equation}

\begin{equation}
q=-\frac{1+z}{2}dlnf(z)/dz.  \label{qz}
\end{equation}%
Now we conclude that if the function $f(z)$ is given, the evolutions of all
the cosmic observable parameters in (\ref{wxz}) - (\ref{qz}) could be
determined uniquely.

\section{The Function $f(z)$ and the Evolutions of Cosmic Parameters}

\begin{figure}[t]
\centerline{\psfig{file=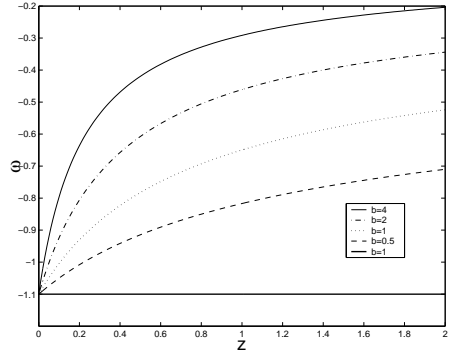,width=6cm}} \vspace*{8pt}
\caption{EOS $w_{x}$ of the dark energy as a function of the redshift $z$
with it's current value $w_{0}=-1.1$ and the bending parameter $b=0$, $1/2$%
, $1$, $2$, $4$, respectivvely.}\label{w_z}
\end{figure}

The parameterization of EOS of the dark energy given by Wetterich has been
extensively studied\cite{EOS Wett,EOS Bartelmann}. It is of the form\cite{Wetterich}%
\begin{equation}
w_{x}(z,b)=\frac{w_{0}}{1+b\ln(1+z)},  \label{EOS}
\end{equation}%
where $w_{x}(z,b)$ is the EOS parameter with it's current value as $w_{0}$,
and $b$ is a bending parameter describing the deviation of $w_{x}$ from $%
w_{0}$ as $z$ increases. Let $w_{0}=-1.1$, we plot the function
(\ref{EOS}) in Fig. \ref{w_z} where $b$ takes the value $0$,
$1/2$, $1$, $2$, $4$, respectively. From this figure we see that
$w_{x}$ varies with $z$ sensitively at low redshift. At high
redshift, it is near to a constant. By
properly choosing the two parameters $w_{0}$ and $b$, the transition from $%
w_{x}<-1$ to $w_{x}>-1$ and the last scattering point at $z\approx 1100$ can
be adjusted easily\ to fit cosmic observations.

Consider equation (\ref{wxz}) now. With use of (\ref{Omegamz}), (\ref%
{Omegarz}) and (\ref{EOS}), we find that (\ref{wxz}) is actually a nonlinear
first-order differential equation. This equation can be integrated out
analytically, giving the solution

\begin{eqnarray}
f(z,b) &=&{(1+z)^{-1}[(1+b\ln(1+z))^{3w_{0}/b}+\Omega
_{m0}-(1+b\ln(1+z))^{3w_{0}/b}\Omega _{m0}+}  \nonumber \\
&&{+\Omega _{r0}+z\Omega _{r0}-(1+b\ln(1+z))^{3w_{0}/b}\Omega
_{r0}]}^{-1}. \label{f(z)}
\end{eqnarray}%
For $b=0$ we have $w(z,0)=w_{0}$. For $b\rightarrow 0$, we find $%
f(z,b)\rightarrow $ $f(z,0)$ with

\begin{eqnarray}
f(z,0) &=&(1+z)^{-1}[(1+z)^{3w_{0}}+\Omega _{m0}-(1+z)^{3w_{0}}\Omega _{m0}+\Omega _{r0}+z\Omega _{r0}-  \nonumber \\
&&{\ -(1+z)^{3w_{0}}\Omega _{r0}]^{-1}.}  \label{f(z0)}
\end{eqnarray}%
Furthermore, for $z=0$ we have $f(0,0)=1$ as it should be by it's definition.

The function $f(z,b)$ shown in (\ref{f(z)}), including the bending parameter
$b$ given in (\ref{EOS}), could, in principle, be determined by
observational data. Be aware that $f\equiv \mu _{0}^{2}/\mu ^{2}$, so we
arrive at a conclusion that the arbitrary function $\mu $ contained in the $%
5D$ solution (\ref{5-metric}) - (\ref{A-B}) could be determined, in terms of
the redshift $z$, by observational data. As for another arbitrary function $%
\nu (z)$, it can be expressed via $\mu (z)$ and $A(z)$ by solving (\ref{A-B}%
) itself. Therefore, $\nu (z)$ is now not arbitrary anymore but is
determined by observational data too. In this way, the whole $5D$ solution
could be determined in principle.

\begin{figure}[t]
\centerline{\psfig{file=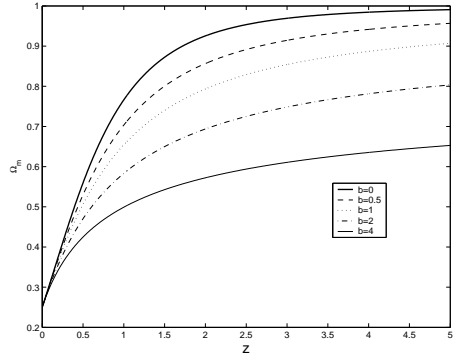,width=6cm}} \vspace*{8pt}
\caption{Evolution of matter density $\Omega _{m}$ versus redshift $z$ with $%
w_{0}=-1.1$, $\Omega _{m0}=0.3$, $\Omega _{r0}=0.00005$, $\Omega _{x0}=0.7$%
. }
\label{Omega_m}
\end{figure}

Return back to (\ref{Omegamz}) - (\ref{qz}). If $f(z,b)$\ is known, all the
three densities $\Omega _{m}$, ${\Omega _{r}}$, $\Omega _{x}$ and the
deceleration parameter $q$ are also known. The evolutions of these three
densities and the deceleration parameter are plotted in Figs.~\ref{Omega_m}-%
\ref{q}.

\begin{figure}[t]
\centerline{\psfig{file=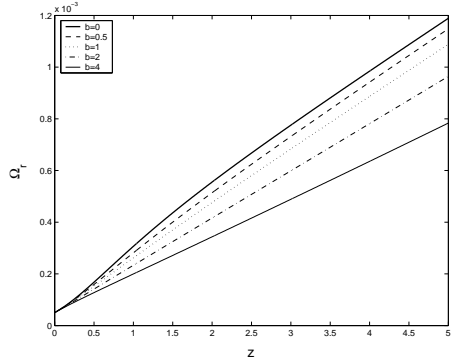,width=6cm}} \vspace*{8pt}
\caption{Evolution of the density $\Omega _{r}$ of radiation
versus redshift $z$ with $w_{0}=-1.1$, $\Omega _{m0}=0.3$, $\Omega
_{r0}=0.00005$, $\Omega _{x0}=0.7$. }
\label{Omega_r}
\end{figure}

\begin{figure}[t]
\centerline{\psfig{file=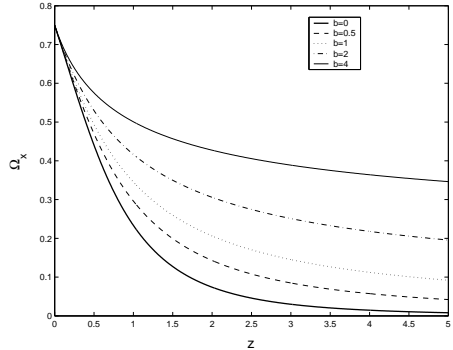,width=6cm}} \vspace*{8pt}
\caption{Evolution of the density $\Omega _{x}$ of dark energy
versus redshift $z$ with $w_{0}=-1.1$, $\Omega _{m0}=0.3$, $\Omega
_{r0}=0.00005$, $\Omega _{x0}=0.7$. }
\label{Omega_x}
\end{figure}

\begin{figure}[t]
\centerline{\psfig{file=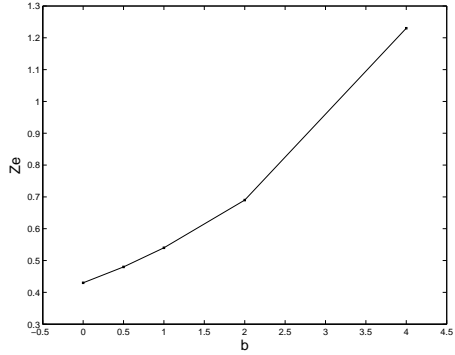,width=6cm}} \vspace*{8pt}
\caption{This figure shows how the bending parameter $b$ affect
the transition point $z_{e}$. The bigger the bending parameter is,
the earlier the transition from mater dominated to dark energy
dominated happend.} \label{ze-b}
\end{figure}

From Fig. \ref{Omega_m}\ to Fig. \ref{Omega_x} we see that as the
redshift $z
$ increases, both $\Omega _{m}$ and ${\Omega _{r}}$ increase too while $%
\Omega _{x}$ decreases, and ${\Omega _{r}}$ increases almost linearly at low
redshift. The effect of the bending parameter $b$ on the three densities is
sensitive. At high redshift it becomes much larger. Clearly, there is a
transition at $z=z_{e}$ at which $\Omega _{m}=\Omega _{x}$. When $z<z_{e}$,
dark energy governs the universe; when $z>z_{e}$, matter becomes the
dominant part of the universe. For several different values of $b$ we plot
the transitions $z_{e}$ in Fig. \ref{ze-b}. If we traced back much earlier
(around $z=5000$), there must be another transition $z=z_{r}$ at which ${%
\Omega _{r}=}\Omega _{m}$ and before which (at $z>z_{r}$) the density ${%
\Omega _{r}}$ dominates the universe. From Fig. \ref{q} we can find how the
bending parameter $b$ affect the deceleration parameter $q$. The transition
from deceleration to acceleration can easily be seen from this figure.
\begin{figure}[t]
\centerline{\psfig{file=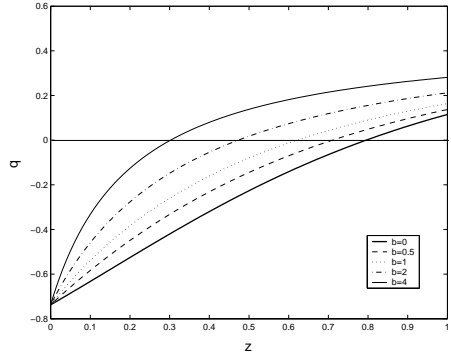,width=6cm}} \vspace*{8pt}
\caption{Evolution of the deceleration parameter $q$ versus
redshift $z$ with $w_{0}=-1.1$, $\Omega _{m0}=0.3$, $\Omega
_{r0}=0.00005$, $\Omega _{x0}=0.7$.} \label{q}
\end{figure}
\section{Conclusions}

The $5D$ Ricci-flat cosmological solution (\ref{5-metric}) - (\ref{A-B})
contains two arbitrary functions $\mu (t)$ and $\nu (t),$ and usually it is
not easy to be determined for a real universe model. In this paper we have
considered the case where the universe is composed of three major
components: matter, radiation, and dark energy, and we have supposed the
equation of state of dark energy is of Wetterich's parameterization form.
Then we show that with use of the relation between the scale factor $A$ and
the redshift $z$, $A=A_{0}/(1+z)$, one can easily change the arbitrary
function $\mu (t)$ to another arbitrary function $f(z)$.\ Furthermore, we
show that this $f(z)$ could be integrated out analytically. Thus, if the
current values of the three density parameters $\Omega _{m0}$, ${\Omega _{r0}%
}$, $\Omega _{x0}$, the EOS $w_{0}$, and the bending parameter $b$ contained
in the EOS are all known, this $f(z)$ could be determined uniquely, and then
both $\mu (z)$ and $\nu (z)$ could be\ determined too. In this way, the
whole $5D$ solution could be constructed and this $5D$ solution could, in
principle, provide with us a global cosmological model to simulate our real
universe. We have also studied the evolutions of the mass density $\Omega
_{m}$, the radiation density ${\Omega _{r}}$, the dark energy density $%
\Omega _{x}$, and the deceleration parameter $q$, and we find that they all
are sensitively dependent on the values of the bending parameter $b$. Thus
we expect that more accurate observational constraints, such as that on the
last-scattering surface and those about the transition points from $\Omega
_{r}$ dominated to $\Omega _{m}$ dominated, from $\Omega _{m}$ dominated to $%
\Omega _{x}$ dominated, and from decelerating expansion to accelerating
expansion of $q$ could help greatly to determine the bending parameter $b$
and then to determine the global evolution of the universe.

\section*{Acknowledgments}

This work was supported by NSF (10573003), NBRP (2003CB716300) of
P. R. China and NSERC of Canada.

\end{document}